\documentclass[a4paper,bibtotoc]{scrartcl}
\usepackage{amsmath}
\usepackage{amssymb}
\usepackage[latin9]{inputenc}
\usepackage{amsthm}
\usepackage{setspace}
\usepackage{longtable}
\usepackage{graphicx}

\begin{document}
\theoremstyle{definition}
\newtheorem{theorem}{Theorem}[section]
\newtheorem{definition}[theorem]{Definition}
\newtheorem{example}[theorem]{Example}
\newcommand{\red}[1]{#1}

\author{Arno Pauly\thanks{Computer Laboratory, University of Cambridge,
  Cambridge CB3 0FD, United Kingdom} \\ \small Arno.Pauly@cl.cam.ac.uk}
\title{The Complexity of Iterated Strategy Elimination}
\maketitle

\begin{abstract}
We consider the computational complexity of the question whether a certain strategy can be removed from a game by means of iterated elimination of dominated strategies. In particular, we study the influence of different definitions of domination and of the number of different payoff values. Additionally, the consequence of restriction to constant-sum games is shown.
\end{abstract}
\section{Introduction}
If I have good reasons not to use a certain option in a strategic situation, or can safely assume that my opponent will not use some of his options, then the situation can be considered as equivalent to a simpler one, in which the respective options are not available. Through the elimination of some of this options, however, it could become possible to discard further options.

The procedure described above serves as a generic template for several solution concepts in game theory, which will now be presented in more detail. In general, we consider two player games in normal form, that is given by a pair $(A, B)$ of $n \times m$ payoff matrices. For a fixed pair of matrices, we define several relations between subgames, that is pairs $(I, J)$ with $I \subseteq \{1, \ldots, n\}$, $J \subseteq \{1, \ldots, m\}$. They differ by the justification needed to delete a certain strategy from the game.
\begin{definition}[Strict Dominance]
The notion of strict dominance is defined through:

\begin{enumerate}
\item $(I, J) \rightarrow_{<<} (I \setminus \{i\}, J)$, if there is an $i_0 \in I$ with $A_{i_0,j} > A_{i, j}$ for all $j \in J$.
\item $(I, J) \rightarrow_{<<} (I, J \setminus \{j\})$, if there is a $j_0 \in J$ with $B_{i,j_0} > B_{i, j}$ for all $i \in I$.
\item $\rightrightarrows_{<<}$ denotes the reflexive and transitive closure of $\rightarrow_{<<}$.
\end{enumerate}
\end{definition}

\begin{definition}[Dominance\footnote{The notion called dominance here, following \cite{kalaic}, is called weak dominance in some parts of the literature.}]
The notion of dominance is defined through:

\begin{enumerate}
\item $(I, J) \rightarrow_{<} (I \setminus \{i\}, J)$, if there is an $i_0 \in I$ with $A_{i_0,j} \geq A_{i, j}$ for all $j \in J$, and a $j_0 \in J$ with $A_{i_0, j_0} > A_{i, j_0}$.
\item $(I, J) \rightarrow_{<} (I, J \setminus \{j\})$, if there is a $j_0 \in J$ with $B_{i,j_0} \geq B_{i, j}$ for all $i \in I$, and an $i_0 \in I$ with $B_{i_0, j_0} > B_{i_0, j}$.
\item $\rightrightarrows_{<}$ denotes the reflexive and transitive closure of $\rightarrow_{<}$.
\end{enumerate}
\end{definition}

\begin{definition}[Weak Dominance\footnote{The notion called weak dominance here, following \cite{kalaic}, is called very weak dominance in some parts of the literature.}]
The notion of weak dominance is defined through:

\begin{enumerate}
\item $(I, J) \rightarrow_{\leq} (I \setminus \{i\}, J)$, if there is an $i_0 \in I$ with $A_{i_0,j} \geq A_{i, j}$ for all $j \in J$.
\item $(I, J) \rightarrow_{\leq} (I, J \setminus \{j\})$, if there is a $j_0 \in J$ with $B_{i,j_0} \geq B_{i, j}$ for all $i \in I$.
\item $\rightrightarrows_{\leq}$ denotes the reflexive and transitive closure of $\rightarrow_{\leq}$.
\end{enumerate}
\end{definition}

If, for some mode of domination, we have $(I, J) \rightarrow (I \setminus \{i\}, J)$ witnessed by $i_0$, we say that $i$ is (weakly / strictly) dominated by $i_0$. Provided that we actually move to the subgame $(I \setminus \{i\}, J)$, we say that $i$ is eliminated by $i_0$. An analogous convention is used for column-player's strategies.

Once a game and a mode of elimination have been specified, the goal is to find a minimal subgame, that is a subgame not further reducible. For all three notions, Nash equilibria of the subgame are also Nash equilibria of the original game, so iterated strategy elimination can be regarded as a pre-processing step in the computation of Nash equilibria. For iterated elimination of strictly dominated strategies the converse is also true: Strategies used in Nash equilibria are never eliminated.

It is known that there is a unique minimal element for strict dominance ($\rightrightarrows_{<<}$) for each game with finite strategy sets, while this is not true for dominance or weak dominance\footnote{The claim of order invariance up to strategy permutation for weak dominance given in \cite[Proposition 1]{knuth} is wrong.}. For weak dominance, uniqueness up to equivalence can be recovered for zero-sum games, or more generally, for games with jointly varying payoffs, as shown in \cite{kalaic}. A game has jointly varying payoffs, if $A_{ij} = A_{kl} \leftrightarrow B_{ij} = B_{kl}$.

Instead of eliminating just one dominated strategy at each step, it is possible to eliminate all currently dominated strategies at once\footnote{Doing so for strictly dominated strategies does not influence the final result, while doing so for weakly dominated strategies could result in empty strategy sets.}. As argued in \cite{gilli}, this notion can be axiomatically justified, and it yields a unique minimal result:
\begin{definition}[Simultaneous Dominance]

\item $(I, J) \rightarrow_{s<} (I \setminus K, J \setminus L)$, if
\begin{enumerate}
\item There are $i_{k} \in I$, $j_{k} \in J$ for each $k \in K$ with $A_{i_k, j} \geq A_{k, j}$ for all $j \in J$ and $A_{i_k, j_k} > A_{k, j_k}$.
\item There are $i_l \in I$, $j_l \in J$ for each $l \in L$ with $B_{i, j_l} \geq B_{i, l}$ for all $i \in I$ and $B_{i_l, j_l} > B_{i_l, l}$.
\item $K \subseteq I$ and $L \subseteq J$ are maximal among all subsets fulfilling 1. and 2.
\end{enumerate}
The transitive and reflexive closure of $\rightarrow_{s<}$ will be denoted by $\rightrightarrows_{s<}$.
\end{definition}

Another notion we will consider is the elimination of never best responses (against pure strategies). This concept belongs to the realm of rationalizability conditions as considered in general in \cite{apt}. The probably best known definitions of rationalizability are the definitions of \cite{pearce} and \cite{bernheim}. Here a strategy will be eliminated, if it is not a best response to any of the remaining strategies for the opponent. As shown in \cite{apt}, there is a unique maximal reduced subgame, provided that one starts with finite strategy sets.
\begin{definition}[Elimination of never best responses]
The notion of iterated elimination of never best responses against pure strategies is defined through:

\begin{enumerate}
\item $(I, J) \rightarrow_{br} (I \setminus \{i\}, J)$, if for each $j \in J$ there is an $i_j \in I$ with $A_{i_j,j} > A_{i, j}$.
\item $(I, J) \rightarrow_{br} (I, J \setminus \{j\})$, if for each $i \in I$ there is a $j_i \in J$ with $B_{i, j_i} > B_{i, j}$.
\item $\rightrightarrows_{br}$ denotes the reflexive and transitive closure of $\rightarrow_{br}$.
\end{enumerate}
\end{definition}

\section{The Computational Problems}
There are several possible ways to obtain computational problems from the notions of iterated elimination of strategies, a variety of them for strict dominance, dominance and weak dominance have been studied in \cite{gilboa3}, showing all of them to be $NP$-complete for dominance and weak dominance, and most of them to be in $P$ for strict dominance. Here we will study problems similar to those considered in \cite{knuth} or \cite{conitzer}, asking whether it is possible to eliminate a certain strategy.

\begin{definition}
$\textsc{Strict}$ has an $n \times m$ game $A$, $B$ and a strategy $1 \leq i \leq n$ as input, and answers $\textsc{yes}$, if there are $I, J$ with $(\{1, \ldots, n\}, \{1, \ldots. m\}) \rightrightarrows_{<<} (I, J)$ and $i \notin I$.
\end{definition}

\begin{definition}
$\textsc{Dominance}$ has an $n \times m$ game $A$, $B$ and a strategy $1 \leq i \leq n$ as input, and answers $\textsc{yes}$, if there are $I, J$ with $(\{1, \ldots, n\}, \{1, \ldots. m\}) \rightrightarrows_{<} (I, J)$ and $i \notin I$.
\end{definition}

\begin{definition}
$\textsc{Weak}$ has an $n \times m$ game $A$, $B$ and a strategy $1 \leq i \leq n$ as input, and answers $\textsc{yes}$, if there are $I, J$ with $(\{1, \ldots, n\}, \{1, \ldots. m\}) \rightrightarrows_{\leq} (I, J)$ and $i \notin I$.
\end{definition}

\begin{definition}
$\textsc{Simultaneous}$ has an $n \times m$ game $A$, $B$ and a strategy $1 \leq i \leq n$ as input, and answers $\textsc{yes}$, if there are $I, J$ with $(\{1, \ldots, n\}, \{1, \ldots. m\}) \rightrightarrows_{s<} (I, J)$ and $i \notin I$.
\end{definition}

\begin{definition}
$\textsc{Response}$ has an $n \times m$ game $A$, $B$ and a strategy $1 \leq i \leq n$ as input, and answers $\textsc{yes}$, if there are $I, J$ with $(\{1, \ldots, n\}, \{1, \ldots. m\}) \rightrightarrows_{br} (I, J)$ and $i \notin I$.
\end{definition}

There are several interesting modifications to the problems introduced above, focusing on additional properties of the game. For $\textsc{Elimination} \in \{\textsc{Strict}, \textsc{Dominance}, \textsc{Weak}, \linebreak \textsc{Simultaneous}, \textsc{Response}\}$, we use $k\textsc{-Elimination}$ to denote the restriction of the respective problem to games with at most $k$ different payoff values, that is $|\{A_{ij} \mid i \leq n, j \leq m\} \cup \{B_{ij} \mid i \leq n, j \leq m\}| \leq k$. $\textsc{Z-Elimination}$ refers to the restriction to zero-sum (or constant sum\footnote{It is straight-forward that these cases are equivalent. Technically, our examples will be constant-sum games rather than zero-sum games.}) games. $k\textsc{-Z-Elimination}$ is defined in the straight forward way.

\section{Previous and new Results}
In each step of any elimination process at least one strategy has to be eliminated, otherwise the elimination stops. Thus, any elimination process is of polynomial length, and can be guessed and verified in polynomial time. Therefore, all problems introduced above are trivially decidable in $NP$. For the notions with unique minimal result, membership in $P$ follows with the same reasoning.

 $NP$-completeness for $2\textsc{-Dominance}$ was established in \cite{conitzer}, which of course implies $NP$-completeness for $k\textsc{-Dominance}$ $(k > 2)$ and for $\textsc{Dominance}$. In \cite{knuth}, $P$-hardness\footnote{While $P$-completeness is claimed, the corresponding part of the proof is wrong. However, as we will show later, the $P$-completeness result is true.} is shown for $6\textsc{-Z-Weak}$. That $\textsc{Z-Dominance}$ can be solved in $P$ was independently shown in \cite{harrenstein}.

 In the present paper, we show that $k\textsc{-Strict}$ is $P$-complete for $k \geq 3$, and $k\textsc{-Z-Strict}$ is $P$-complete for $k \geq 4$. Both $2\textsc{-Strict}$ and $3\textsc{-Z-Strict}$ are $NL$-complete. In a zero-sum game with payoffs in $\{0, 1\}$, only trivial cases of strict dominance are possible, and there cannot be any iteration of elimination. This allows to decide $2\textsc{-Z-Strict}$ in $L$.

 By proving them to be equivalent to $2\textsc{-Strict}$, also $\textsc{Response}$ and $k\textsc{-Response}$ $(k \geq 2)$ will be shown to be $NL$-complete.

 We establish (again) polynomial time decidability of $\textsc{Z-Dominance}$. For $3\textsc{-Z-Dominance}$ as well as for $3\textsc{-Z-Simultaneous}$ we show $P$-completeness, leaving the case $k = 2$ open. Without the restriction to zero-sum games, we can show that $2\textsc{-Simultaneous}$ is $P$-complete.

 For weak dominance, we establish the membership in $P$ of $\textsc{Z-Weak}$, and reduce the number of payoff values needed for $P$-completeness, so that we can prove $3\textsc{-Z-Weak}$ to be $P$-complete. The problem $3\textsc{-Weak}$ is $NP$-complete. For the case of just two different payoff values, we only establish $P$-hardness of $2\textsc{-Weak}$, and leave the remaining questions open.

\section{Complexity classes and there complete problems}
In this section, we shall briefly introduce the complexity classes occurring in our results, and present their complete problems we use to derive the completeness-results. Most of this section is based on \cite{papadimitrioud}.

The class $NP$ captures the problems decidable by a polynomially time-bounded nondeterministic Turing machine. $NP$-complete problems are not assumed to admit fast decision algorithms. A prototypic $NP$-complete problem is \textsc{3SAT}, defined as following:
\begin{definition}[3-Satisfiability]
\textsc{3SAT} has a list of clauses as input, each containing three literals of the form $X_i$ or $\neg X_i$. The question is whether truth values can be attached to the literals, so that at least one literal per clause is true.
\end{definition}

A polynomially time-bounded deterministic Turing machine decides the problems in $P$. The concept of $P$-completeness captures the problems where a substantial speed-up through parallel computing is least to be expected, i.e. the inherently sequential problems. As a prototypic $P$-complete problem we use several versions of the monotone circuit value problem ($\textsc{MCV}$):
\begin{definition}[monotone circuit value problem]
The problem $\textsc{MCV}$ takes a monotone circuit as input, that is a directed acyclic graph ($\textsc{Dag}$), where the vertices are labeled with $\textsc{And}$, $\textsc{Or}$ and $\textsc{False}$. Vertices labeled with $\textsc{False}$ always have in-degree 0, while $\textsc{And}$ and $\textsc{Or}$ vertices have in-degree less or equal 2. The value of a \textsc{False} vertex is \emph{false}. An \textsc{And} vertex has the value \emph{true}, if and only if all his input vertices have value \emph{true}\footnote{In particular, an \textsc{And} vertex with in-degree 0 always has the value \emph{true}; thus, we do not need to include designated \textsc{True} vertices. In theory, the same is true for \textsc{False} and \textsc{Or} vertices, however, including \textsc{False} vertices explicitly facilitates our constructions.}; an \textsc{Or} vertex has value \emph{true}, if and only if he has an input vertex with value \emph{true}. There is exactly one vertex with out-degree 0, the root. The answer to the problem is $\emph{yes}$, if and only if the root is assigned the value $\emph{true}$.
\end{definition}

There are two different sets of additional restrictions imposed on the circuits we will use, both lead to problems equivalent to the original $\textsc{MCV}$:

\begin{definition}
In the problem $\textsc{MCV1}$, $\textsc{And}$ and $\textsc{Or}$ vertices are alternating. Each $\textsc{False}$ vertex is input to a specific $\textsc{Or}$ vertex; \textsc{And} vertices only have $\textsc{Or}$ vertices as input (if any). All $\textsc{Or}$ vertices have in-degree 2. Two vertices never share all their inputs (except when the set is empty). The root is labelled \textsc{And}. There is at least one vertex with each label.
\end{definition}

\begin{definition}
In the problem $\textsc{MCV2}$, $\textsc{And}$ and $\textsc{Or}$ vertices are alternating. \textsc{Or} vertices only have $\textsc{And}$ vertices as input. All $\textsc{Or}$ vertices have in-degree 2. Each $\textsc{And}$-vertex has at most one $\textsc{False}$-vertex as input. The root is labelled \textsc{And}, and has no $\textsc{False}$-vertex as input. Different $\textsc{And}$-vertices have disjoint inputs, different $\textsc{Or}$-vertices have unequal inputs.
\end{definition}

The third complexity class needed is $NL$, the class of problems decidable on a logarithmically space-bounded nondeterministic Turing machine. The standard complete problem for $NL$ is \textsc{Reachability}, defined as:

\begin{definition}[Reachability]
The problem \textsc{Reachability} takes a \textsc{Dag} $G$ together with two vertices $s$, $t$ as input, and answers \textsc{Yes}, if there is a path in $G$ from $s$ to $t$.
\end{definition}

For our purposes, another problem is more useful:

\begin{definition}[Cycle Reachability]
The problem $\textsc{CycleReach}$ takes a directed graph $G$ and a vertex $s$ as input, and answers \textsc{Yes}, if $G$ contains a cycle which can be reached from $s$.
\end{definition}

\begin{theorem}
$\textsc{CycleReach}$ is $NL$-complete\footnote{I would like to thank Anuj Dawar and Yuguo He for pointing out this result to me.}.
\begin{proof}
A non-deterministic algorithm for $\textsc{CycleReach}$ storing only a constant number of vertices works as follows. Guess a node $u \in V$ which is kept for the rest of the algorithm. Starting with $v$ as the active vertex, always guess a successor of the active vertex, and let it be the new active vertex. If there is no successor, reject. If the active vertex equals $u$ for the first time, flip a control bit. If it equals $u$ for the second time, accept. In addition, the number of steps can be counted, and the computation can be aborted once its exceeds $2|V|$.

To see that $\textsc{CycleReach}$ is even $NL$-complete, we present a reduction from $\textsc{Reachability}$. If a path from $s$ to $t$ is sought, an edge from $t$ to $s$ is added. Then a cycle can be reached from $s$, iff $t$ was reachable from $s$ in the original graph.
\end{proof}
\end{theorem}

\section{Inside $NL$}
\begin{theorem}
$2\textsc{-Z-Strict}$ is in $L$.
\begin{proof}
The only possible dominations are by a row of $1$s against a row of $0$s, or by a column of $0$s against a column of $1$s. If we assume that the initial game allows to eliminate a row, there must be a row containing only $1$s, which of course is uneliminable. Thus, there will never be a column containing only $0$s, that means there will never be a domination between columns.

The considerations above show that the following algorithm is sufficient to solve $2\textsc{-Z-Strict}$. In the first step, determine whether the game has a row containing only $1$s, or a column containing only $0$s, or neither. In the first case, the initial game is copied to the output tape row-wise, leaving out all $0$-rows, in the second case, it is copied column-wise, leaving out all $1$-columns. In the third case, the complete game forms the output.
\end{proof}
\end{theorem}

\begin{theorem}
$2\textsc{-Strict}$ is $NL$-complete.
\label{theorem2strict}
\begin{proof}
We show that $2\textsc{-Strict}$ and $\textsc{CycleReach}$ are equivalent. 

Given a game, we can check in logarithmic space whether there are strategies for both row and column player always granting a payoff of $1$ to the respective player after the first round of elimination. If not, the iteration stops and the answer can be determined already. If there are such strategies, we consider the remaining strategies of both players as vertices in a graph. There is an edge from $s_{i}$ to $t_{j}$, iff $A_{ij} = 1$, and an edge from $t_{j}$ to $s_i$ iff $B_{ij} = 1$. There are no edges between $s_{i}$ and $s_{k}$ or between $t_{j}$ and $t_l$. Iterated elimination of strictly dominated strategies now corresponds to iteratively removing vertices without outgoing edges. In the end, only those vertices remain from which a cycle can be reached.

For the other direction, we start with transforming the graph into a bipartite graph by inserting a new vertex for each edge. Then we construct a game, where both player have a special strategy always yielding payoff $1$, and a strategy for each vertex in their set of vertices. An edge corresponds to payoff $1$, no edge to payoff $0$. Again, removal of dominated strategies corresponds to removal of vertices without outgoing edges.
\end{proof}
\end{theorem}

\begin{theorem}
$3\textsc{-Z-Strict}$ is $NL$-complete.
\begin{proof}
We start with presenting a reduction from $\textsc{CycleReach}$ to the problem at hand. We can assume the graph to be bipartite without influencing the complexity of the problem. In addition, we can assume that the graph does not contain a cycle of length $2$, either by inserting additional edges, or by testing in logarithmic space. The game to be constructed has fixed strategies $s$ and $t$ for row or column players respectively, and additional strategies $s_u$ and $t_v$ for nodes $u \in V_1$ or $v \in V_2$, if $\{V_1, V_2\}$ is the partition of the vertex set. The payoffs for row player are as follows:
$$\begin{array}{c|cc}
& t & t_v \\
\hline
s & 1 & 2 \\
s_u & 0 & \begin{cases} 2 & (u, v) \in E \\ 0 & (v, u) \in E \\ 1 & \textnormal{otherwise}\end{cases}
\end{array}$$
A strategy $t_v$ gets eliminated by $t$ iff $v$ has no successors in the graph, the same holds for $s_u$. As $s$ and $t$ are never eliminated, there cannot be eliminations of $t_{v_1}$ by $t_{v_2}$ or of $s_{u_1}$ by $s_{u_2}$.

Now we have to show that iterated elimination can be executed in $NL$ under the current restrictions. As all instances where there is a row consisting only of $2$s or a column consisting only of $0$s are trivial, we note the following properties for games containing neither:
\begin{enumerate}
\item A row can be dominated only if it contains a $0$.
\item A row can dominate another row only if it does not contain a $0$.
\item A row cannot be dominated if it contains a $2$.
\item A column can be dominated only if it contains a $2$.
\item A column can dominate another row only if it does not contain a $2$.
\item A column cannot be dominated if it contains a $0$.
\end{enumerate}
Combining these observations, an elimination never changes whether a row contains $0$s or not, or whether a column contains $2$s or not. Thus, each row (column) is potentially dominating, that is contains only $1$s and $2$s ($1$s and $0$s) or not. We use $R_D$ ($C_D$) to denote the set of potentially dominating rows. If some column $c$ returns $0$ or $1$ against any row in $R_D$, then $c$ cannot be eliminated, since no column can return strictly less against the same row, and the row is uneliminable. By the same argument, each potentially eliminable row must return $0$ against any column from $C_D$, the corresponding sets shall be called $R_E$ and $C_E$.

The sets $R_E$ and $C_E$ are now used as vertices in a graph: There is an edge from $v \in R_E$ to $u \in C_E$, if $v$ against $u$ yields $2$ and an edge from $u \in C_E$ to $v \in R_E$, if $u$ against $v$ returns $0$. If a vertex has a successor in this graph, it cannot currently be eliminated. However, the converse is not necessarily true.

Therefore, we have to test for each $v \in R_E$ whether there is a $\hat{v} \in R_D$, so that $\hat{v}$ yields strictly more than $v$ against all $u \notin C_E$. If there is no such strategy, we add a cycle and an edge from $v$ to the cycle to our graph. The same procedure is executed for columns. The existence of successors still implies non-eliminability.

If a row-vertex $v$ has no successor, then there is a row $\hat{v} \in R_D$ yielding better rewards against all $u \notin C_E$. By definition, $\hat{v}$ yields $2$ against all $u \in C_E$, while $v$ yields less. Thus, $\hat{v}$ strictly dominates $v$, and $v$ can be eliminated. The same holds for columns, showing that we have presented a reduction to $\textsc{CycleReach}$.
\end{proof}
\end{theorem}

\begin{theorem}
$k\textsc{-Response}$ is $NL$-complete for $k \geq 2$. $\textsc{Response}$ is $NL$-complete.
\begin{proof}
Considering the trivial reducibilities between the concerned problems, it is sufficient to show that $2\textsc{-Reponse}$ is $NL$-hard and that $\textsc{Response}$ is in $NL$. For the former, note that the reduction from $\textsc{CycleReach}$ to $2\textsc{-Strict}$ presented in the proof of Theorem \ref{theorem2strict} also is a reduction from $\textsc{CycleReach}$ to $2\textsc{-Response}$.

To show membership in $NL$ for $\textsc{Response}$, we present a reduction to $2\textsc{-Strict}$. Given the payoff matrix $A$ for row-player, we construct a new matrix $\hat{A}$, where $\hat{A}_{ij} = 1$, iff $A_{ij} \geq A_{ik}$ for all $k$. For column player, the best payoffs in each row are replaced by $1$, all other values by $0$. In the resulting game, the best-response-relationships are unchanged. By adding a new strategy always returning $1$ for each player, the best-response-relations of the previously present strategies are still unchanged. Never-best-responses are strategies always yielding $0$, and these strategies will be dominated by the all $1$-strategy.
\end{proof}
\end{theorem}

\section{$P$-completeness}
We start with presenting polynomial time algorithms for the problems admitting one (regardless of the number of different payoff values), and then show hardness for the lowest $k$-value possible. Note that the following algorithm can be executed in polynomial time: \emph{Search for a (weakly / strictly) dominated strategy. If there is one, eliminate it and start again. If there is none, check whether the specified strategy is still there or not.} Provided that the order of elimination is irrelevant, this algorithm solves the problems at hand. This directly leads to:
\begin{theorem}
$\textsc{Strict}$ is in $P$.
\end{theorem}

\begin{theorem}
\label{simultaneousinp}
$\textsc{Simultaneous}$ is in $P$.
\begin{proof}
Only switch from elimination of rows to elimination of columns and vice versa if necessary.
\end{proof}
\end{theorem}

The remaining cases require more work. Basically, we will show in both cases that a certain order of elimination is sufficient to eliminate everything that could be eliminated.

\begin{theorem}
$\textsc{Z-Weak}$ is in $P$.
\begin{proof}
When the naive approach finds a way to eliminate the strategy $s$, then it will yield the right answer. The only problem occurs if the initial game $(A, B)$ is reduced to a game $(\hat{A}, \hat{B})$, so that $s$ is iteratively eliminable in $(A, B)$, but not in $(\hat{A}, \hat{B})$. So we assume that there is a sequence of eliminations $(x_i, y_i)$, so that in the $i$th step the strategy $x_i$ weakly dominates $y_i$, and by that eliminating the latter; the sequence shall end with $y_{i_{\textnormal{max}}} = s$. The case we have to consider is another possible elimination $(x, y)$, that makes one of the eliminations in our sequence impossible. We denote the first elimination being made impossible by $(x_k, y_k)$.

The only way that $(x_k, y_k)$ is made impossible is by elimination of $x_k$, so we know $y = x_k$. If $x \neq y_k$, then $(x_k, y_k)$ could simply be replaced by $(x, y_k)$\footnote{Or, if $x$ is subsequently eliminated by another strategy $z$, by $(z, y_k)$, and so on.}. Thus, only the situation $x_k = y$, $y_k = x$ is problematic. In this case, however, $x_k$ and $y_k$ have to be identical once the $k$th stage has been reached. Thus, elimination of $y_k$ alone does not enable new weak dominations in later stages. Therefore, the only problematic case is $y_k = s$. Clearly, this can be avoided if we never use $s$ as a weakly dominating strategy for eliminations.

The question that remains it whether it might be necessary to eliminate a strategy \emph{by} $s$ to allow later elimination \emph{of} $s$. This means the following situation:
$$\begin{array}{c|ccc} & u & v & \ldots \\
\hline
s & a & b & \ldots \\
r & c & d & \ldots \\
t & e & f & \ldots \\
\ldots & \ldots & \ldots & \ldots
\end{array}$$
\begin{enumerate}
\item $s$ weakly dominates $r$: $a \geq c$, $b \geq d$
\item $u$ does not weakly dominate $v$, but will do so once $r$ has been eliminated: $a \leq b$, $c > d$, $e \leq f$
\item $t$ does not weakly dominate $s$, but will do so once $v$ has been eliminated: $e \geq a$, $f < b$
\item $t$ does not weakly dominate $r$: $e < c \ \vee \ f < d$
\end{enumerate}
We have $f \geq e \geq a \geq c > d$, rendering both $f < d$ and $e < c$ impossible. Thus, $t$ could be used to eliminate $r$ in such a situation. For more steps between the elimination of $r$ and the elimination of $s$, the same considerations apply; showing that elimination by $s$ is never necessary to allow elimination of $s$.

Therefore, the following modification of the algorithm above solves $\textsc{Z-Strict}$: \emph{Search for a strategy weakly dominated by another strategy not equal to the specified strategy. If there is one, eliminate it and start again. If there is none, check whether the specified strategy is still there or not.}
\end{proof}
\end{theorem}

\begin{theorem}
$\textsc{Z-Dominance}$ is in $P$.
\begin{proof}
We assume that $(x_k, y_k)$ is a sequence of eliminations, stating that $y_k$ is eliminated in the $k$th step, which is justified by domination by $x_k$, so that in the last step $s$ is eliminated. Let $(x, y)$ be another elimination, whose execution leads to $(x_l, y_l)$ being the first elimination in the sequence that is impossible. In the case $y = x_l$, we can simply replace $(x_l, y_l)$ by $(x, y_l)$ without substantial changes. The other case of $(x, y)$ making $(x_l, y_l)$ impossible is $y$ being the only strategy against which $x_l$ yields better reward than $y_l$. But then only dominations including $y$ are affected by $(x_l, y_l)$. If there were an $k > l$ with $y_k = y$, this does not matter, since $y$ is already gone. In the case $x_k = y$, the elimination $(x, y_k)$ can be used instead. Therefore, the impossibility of executing $(x_l, y_l)$ has no substantial impact on later eliminations. The only problematic case is $y_l = s$.

As every column strategy could potentially be the strategy providing strictly better payoff in the domination of $s$, the following algorithm is able to solve $\textsc{Z-Dominance}$:
\emph{Pick a column strategy $x$. Search for eliminable strategies unequal to $x$, remove them if found. If there are none left, check whether $s$ has been eliminated. If $s$ is still present, choose another column strategy for $x$ and repeat. If all columns have been tried without $s$ being eliminated, $s$ is uneliminable.}
\end{proof}
\end{theorem}

\begin{theorem}
\label{phardness1}
$3\textsc{-Z-Weak}$, $3\textsc{-Z-Dominance}$ and $3\textsc{-Z-Simultaneous}$ are $P$-hard.
\begin{proof}
The claims will be proven using a reduction from \textsc{MCV1}. From such a circuit, we will construct a zero-sum game. Row-player has a strategy $s_{\wedge n}$ for each \textsc{And} vertex with number $n$, a strategy $s_{\bot m}$ for each \textsc{False} vertex $m$ and another strategy $s_B$. Column-player has the strategies $t_{\wedge i}$, $t_{\vee j}$ and $t_{\bot k}$ for \textsc{And} vertices $\wedge_i$, \textsc{Or} vertices $\vee_j$ and \textsc{False} vertices $\bot_k$.

$$\begin{array}{c|ccc}
 & t_{\wedge i} & t_{\bot k} & t_{\vee j} \\
 \hline
      s_B        & 0 & 0 & 0 \\
 s_{\wedge n}  & \begin{cases} -1 & \textnormal{if } i = n \\ 0 & \textnormal{otherwise }\end{cases} & 0 & \begin{cases} 1 & \textnormal{if $\vee_j$ is an input for $\wedge_n$} \\ -1 & \textnormal{if $\wedge_n$ is an input for $\vee_j$} \\ 0 & \textnormal{otherwise }\end{cases} \\
    s_{\bot m} & 1 & \begin{cases} -1 & \textnormal{if } k = m \\ 1 & \textnormal{otherwise }\end{cases} & \begin{cases} -1 & \textnormal{if $\bot_m$ is an input for $\vee_j$} \\ 0 & \textnormal{otherwise }\end{cases}\\
\end{array}$$

The payoffs of the row-player are given by the table above, the column-player just tries to minimize the payoffs. The elimination of a strategy $s_{\wedge n}$ (and $t_{\wedge n}$) or $t_{\vee j}$ corresponds to assigning the value \emph{true} to the corresponding vertices $n$ or $j$. We will now study which strategies can be removed under which conditions, and see that the strategies $s_B$, $s_{\bot m}$ and $t_{\bot k}$ could only be eliminated, if all strategies $s_{\wedge n}$ have been eliminated first, in which case the answer is already determined anyway.
\begin{description}
\item[$s_B$] The strategy $s_B$ can be weakly dominated by a strategy $s_{\wedge n}$ only if $t_{\wedge n}$ has been eliminated. In this case, $s_{\wedge n}$ was eliminated first. $s_B$ can be weakly dominated by a strategy $s_{\bot m}$ only if $t_{\bot m}$ was eliminated first.

\item[$s_{\wedge n}$] The strategy $s_{\wedge n}$ is weakly dominated by $s_B$, if all strategies $t_{\vee j}$ where $\vee_j$ is an input to $\wedge_n$ have been eliminated. Provided that $t_{\wedge n}$ has not been eliminated yet (the contrary would be impossible), in this case $s_{\wedge n}$ is also dominated by $s_B$. $s_{\wedge n}$ cannot be weakly dominated by a strategy $s_{\wedge n'}$, as long as $t_{\wedge n'}$ is still present, and it cannot be weakly dominated by a strategy $s_{\bot m}$, as long as $t_{\bot m}$ is still present.

\item[$s_{\bot m}$] The strategy $s_{\bot m}$ cannot be weakly dominated by $s_{\wedge n}$, as long as $t_{\wedge n}$ is still present. It cannot weakly dominated by $s_B$, as long as a strategy $t_{\wedge i}$ is present. Weak domination by a strategy $s_{\bot m'}$ is impossible as long as $t_{\bot m'}$ is present.

\item[$t_{\wedge i}$] The strategy $t_{\wedge i}$ cannot be weakly dominated by another strategy $t_{\wedge i'}$ or by a strategy $t_{\bot k}$, as long as $s_{\wedge i}$ is present. If $s_{\wedge i}$ is eliminated, $t_{\wedge i}$ is weakly dominated by any $t_{\bot k}$. $t_{\wedge i}$ cannot be weakly dominated by a strategy $t_{\vee j}$, as long as there is a strategy $s_{\wedge n}$, such that the vertex $\vee_j$ is an input to $\wedge_n$. The existence of such a vertex can be assumed, since the root is labelled $\textsc{And}$, and cannot be eliminated unless all inputs are eliminated first, this cannot happen.

\item[$t_{\bot k}$] The strategy $t_{\bot k}$ cannot be eliminated by a strategy $t_{\wedge i}$ or $t_{\bot k'}$, as long as $s_{\bot k}$ is still present. Weak dominance by $t_{\vee j}$ would only be possible, if no $s_{\wedge n}$ with $\vee_j$ being an input to $\wedge_n$ would exists, as explained above, this does not happen.

\item[$t_{\vee j}$] The strategy $t_{\vee j}$ is dominated by $t_{\wedge i}$ or $t_{\bot k}$, if $\wedge_i$ or $\bot_k$ are the only remaining input to $\vee_j$, which requires the second input to be true. $t_{\vee j}$ could only be eliminated by $t_{\vee j'}$, if $\vee_j$ and $\vee_{j'}$ had the same inputs, which was ruled out in our convention, or if $t_{\vee j}$ could also have been eliminated by certain $t_{\wedge i}$ or $t_{\bot k}$.
\end{description}

\end{proof}
\end{theorem}

\begin{theorem}
$3\textsc{-Strict}$ is $P$-hard.
\begin{proof}
Again a reduction from $\textsc{MCV1}$ is given. The players have the same strategies as in the proof of Theorem \ref{phardness1}, except for $s_B$, which is no longer needed. The payoffs are given by the following tables, with row-players payoffs first:
$$\begin{array}{c|ccc}
 & t_{\wedge i} & t_{\bot k} & t_{\vee j} \\
 \hline
 s_{\wedge n}  & 0 & 0 & \begin{cases} 1 & \textnormal{if $\vee_j$ is an input for $\wedge_n$} \\ 0 & \textnormal{otherwise }\end{cases} \\
    s_{\bot m} & 1 & 1 & 1\\
\end{array}$$ $$
\begin{array}{c|ccc}
 & t_{\wedge i} & t_{\bot k} & t_{\vee j} \\
 \hline
 s_{\wedge n}  & \begin{cases} 1 & \textnormal{if } i = n \\ 0 & \textnormal{otherwise }\end{cases} & 0 & \begin{cases} 0 & \textnormal{if $\wedge_n$ is an input for $\vee_j$} \\ -1 & \textnormal{otherwise }\end{cases} \\
    s_{\bot m} & 0 & \begin{cases} 1 & \textnormal{if } k = m \\ 0 & \textnormal{otherwise }\end{cases} & \begin{cases} 0 & \textnormal{if $\bot_m$ is an input for $\vee_j$} \\ -1 & \textnormal{otherwise }\end{cases}\\
\end{array}$$
It is trivial to see that $s_{\bot m}$ can never be eliminated, and that a strategy $s_{\wedge n}$ is strictly dominated by any strategy $s_{\bot m}$, as soon as all strategies $t_{\vee j}$ corresponding to its input vertices have been removed. Thus, the removal of row-players strategies corresponds exactly to the corresponding vertices being assigned the value \emph{true}.

For column player, a strategy $t_{\wedge i}$ can never be strictly dominated as long as any strategy $s_{\wedge n}$ is still present. As the strategy $s_{\bot m}$ will never be eliminated, elimination of $t_{\bot m}$ is also impossible. The strategy $t_{\vee j}$ can be eliminated by any strategy, if there are no strategies $s_{\wedge n}$ or $s_{\bot m}$ corresponding to input vertices left, but it can also be eliminated by the strategy $t_{\wedge i}$ or $t_{\bot k}$, where $\wedge_i$ or $\bot_k$ is the only remaining strategy corresponding to an input vertex of $\vee_j$ for which $s_{\wedge i}$ or $s_{\bot k}$ is still present. Thus, for $t_{\vee j}$ to be removed, at least one of its input vertices needs to be removed earlier.
\end{proof}
\end{theorem}

\begin{theorem}
$4\textsc{-Z-Strict}$ is $P$-hard.
\begin{proof}
Once more a reduction from $\textsc{MCV1}$ is used; in addition to the conditions listed there, we assume that there are at least 2 $\textsc{Or}$ vertices. Row-player has a fixed strategy $s_B$, a strategy $s_{\wedge i}$ for each $\textsc{And}$-vertex $\wedge_i$ and a strategy $s_{\vee n}$ for each $\textsc{Or}$-vertex $\vee_n$. Column-player has strategies $t_{\vee j}$, $t_{\vee j-1}$ and $t_{\vee j-2}$ for each $\textsc{Or}$-vertex $\vee_j$, where $t_{\vee j-x}$ is only present iff the $x$th input to $\vee_j$ is an $\textsc{And}$-vertex. The payoffs are given by the following table:
$$\begin{array}{c|ccc}
 & t_{\vee j} & t_{\vee j-x} \\
 \hline
  s_B      & 3 & 2\\
  s_{\wedge i}  & \begin{cases} 3 & \textnormal{if $\vee_j$ is an input for $\wedge_i$} \\ 1 & \textnormal{otherwise }\end{cases} & \begin{cases} 1 & \textnormal{if $\wedge_ii$ is the $x$th input to $\vee_j$} \\ 1  & \textnormal{if $\vee_j$ is an input to $\wedge_i$} \\ 0 & \textnormal{otherwise}\end{cases}  \\

 s_{\vee_n} & \begin{cases} 2 & n = j \\ 3 & \textnormal{otherwise}\end{cases} & \begin{cases} 1 & n = j \\ 2 & \textnormal{otherwise} \end{cases}
\end{array}$$

The only strategies that are potentially eliminable are $s_{\wedge_i}$ and $t_{\vee_j}$, corresponding to the respective vertices being assigned the value \emph{true}. As explained below, the other strategies are uneliminable.

\begin{description}
\item[$s_B$] It is obvious that $s_B$ can never be eliminated, since it is a best response against any of column-player's strategies.

\item[$s_{\wedge i}$] If all strategies $t_{\vee j}$ where $\vee_j$ is an input vertex to $\wedge_i$ have been eliminated, the strategy $s_{\wedge i}$ will also be eliminated. If any of these strategies is still present, $s_{\wedge i}$ cannot be removed, since it achieves the maximal payoff against it.

\item[$s_{\vee n}$] The strategies $s_{\vee n}$ are uneliminable. Due to convention, there is an $\textsc{Or}$-vertex $\vee_j$ with $n \neq j$. If the vertex $\vee_j$ has at least one $\textsc{And}$-vertex as input, there is an (uneliminable) strategy $t_{\vee j-x}$ against which $s_{\vee n}$ yields $2$, which cannot be exceeded. If both inputs of $\vee_j$ are $\textsc{False}$-vertices, $t_{\vee j}$ itself is uneliminable, and takes the place of $t_{\vee j-x}$ in the argument above.

\item[$t_{\vee j}$] As $s_{\vee j}$ cannot be eliminated, a strategy $t_{\vee j}$ can only be eliminated by a strategy $t_{\vee j-x}$. This happens, if and only if the strategy $s_{\wedge i}$ corresponding to the $\textsc{And}$-vertex forming the $x$th input of $\vee_j$ was eliminated previously.

\item[$t_{\vee j-x}$] The strategies $t_{\vee j-x}$ can never be strictly dominated in a subgame where $s_B$ is still present. Since $s_B$ can never be eliminated, the same is true for all strategies $t_{\vee j-x}$.
\end{description}
\end{proof}
\end{theorem}

\begin{theorem}
$2\textsc{-Weak}$, $2\textsc{-Dominance}$ and $2\textsc{-Simultaneuous}$ are $P$-hard.
\begin{proof}
This time, we present a reduction from \textsc{MCV2}. The reduction works for all three problems simultaneously, as they yield identical answers for the constructed game.

Row player has a strategy $s_{\wedge i}$ for each \textsc{And} vertex $\wedge_i$, and a strategy $s_B$. Column player has a strategy $t_{\vee j}$ for each $\textsc{Or}$ vertex $\vee_j$, and strategies $t_{\wedge \bot k}$, where $k$ simultaneously enumerates $\textsc{And}$ and $\textsc{False}$ vertices. The $\textsc{And}$-vertex $k$ is referring to is $\wedge_k$, the corresponding $\textsc{False}$-vertex is $\bot_k$. We require $\bot_k$ to be an input of $\wedge_k$, thus, if the root is $\wedge_r$, $\bot_r$ does not exist. The payoffs are given by the following tables, starting with row player:
$$\begin{array}{c|cc}
 & t_{\vee j} & t_{\wedge \bot k} \\
 \hline
 s_{\wedge i} & \begin{cases} 1 & \textnormal{ if $\vee_j$ is input to $\wedge_i$} \\ 0 & \textnormal{else}\end{cases} & \begin{cases} 1 & \textnormal{if $\bot_k$ is input to $\wedge_i$} \\ 0 & \textnormal{else} \end{cases} \\
 s_B & 0 & \begin{cases} 1 & \textnormal{if $\bot_k$ does not exists} \\ 0 & \textnormal{else}\end{cases}
 \end{array}$$

 $$\begin{array}{c|cc}
 & t_{\vee j} & t_{\wedge \bot k} \\
 \hline
 s_{\wedge i} & \begin{cases} 1 & \textnormal{ if $\wedge_i$ is input to $\vee_j$} \\ 0 & \textnormal{else}\end{cases} & \begin{cases} 1 & i = k \\ 0 & \textnormal{else} \end{cases} \\
 s_B & 0 & 1
 \end{array}$$

 For the following considerations, we always assume that the strategy corresponding to the root of the circuit has not been eliminated yet. Otherwise, the answer is already determined, and further elimination do not matter.
\begin{description}
\item[$s_{\wedge i}$] The strategy $s_{\wedge i}$ is weakly dominated by a strategy $s_{\wedge i'}$, if and only if the inputs of $\wedge_i$ are a subset of the inputs of $\wedge_{i'}$, which can only happen if the input set of $s_{\wedge i}$ is empty. If $\wedge_i$ has a $\textsc{False}$-vertex as input, this is $\bot_i$ as input. Then $s_{\wedge i}$ cannot be eliminated by $s_B$, as long as $t_{\wedge \bot i}$ is still present. Since $t_{\wedge \bot i}$ cannot be eliminated as long as $s_{\wedge i}$ is present, this renders $s_{\wedge i}$ uneliminable.

     If all existing inputs of $\wedge_i$ are $\textsc{Or}$-vertices, then $s_{\wedge i}$ is weakly dominated by $s_B$, once all strategies $t_{\vee j}$ corresponding to inputs of $\wedge_i$ have been removed. Since there is a strategy $t_{\wedge \bot k}$, where $\bot_k$ does not exist, in the case of weak dominance, we also have dominance. Thus, elimination of $s_{\wedge i}$ corresponds to the vertex $\wedge_i$ being assigned the value \emph{true}.

\item[$s_B$] By convention, for the strategy $t_{\wedge \bot r}$ corresponding to the root, $\bot_r$ does not exist. As we assume this strategy not to be elimianted yet, and $s_B$ is the only strategy achieving payoff $1$ against it (since a non-existing vertex cannot be the input to another vertex). Thus, $s_B$ is uneliminable.

\item[$t_{\vee j}$] Due to our convention, each vertex $\vee_j$ has exactly two $\textsc{And}$-vertices as input. If the strategy $s_{\wedge i}$ corresponding to one of them is eliminated, then also the strategy $t_{\vee j}$ is eliminated, e.g. by $t_{\wedge \bot k}$, where $\wedge_k$ is a remaining input, or by any other strategy, if there is no remaining input. Due to the strategy $s_B$, weak dominance will already imply dominance.

    If both of the strategies $s_{\wedge i}$, $s_{\wedge i'}$ corresponding to inputs of $\vee_j$ remain, then $t_{\vee j}$ cannot be eliminated. Thus, elimination of $t_{\wedge j}$ corresponds to assigning the value $\emph{true}$ to the vertex $\wedge_j$.

\item[$t_{\wedge \bot k}$] As long as $s_B$ is not eliminated, a strategy $t_{\wedge \bot k}$ can never be eliminated by a strategy $t_{\vee j}$. The strategy $t_{\wedge \bot k}$ might be eliminated by another strategy $t_{\wedge \bot k'}$, only once $s_{\wedge k}$ is eliminated first. In this case, the strategy $t_{\wedge \bot k}$ has no further relevance anyway.
\end{description}
\end{proof}
\end{theorem}

\section{$NP$-completeness}
\begin{theorem}
\label{weakgeneral}
$3\textsc{-Weak}$ is $NP$-complete.
\begin{proof}
A reduction from $3SAT$ shall be presented. We assume that for different clauses $c$ and $c'$, the set of literals occurring in $c$ is never a subset of the set of literals occurring in $c'$. For a clause $c$, $c_i$ refers to the $i$th literal in $c$.

Row-player has a strategy $s$, as well as strategies $s_d$ for each clause $d$ and $s_{l+}$ and $s_{l-}$ for each variable $l$. Column-player has a strategies $t_c$, $t_c^i$ for each clause $c$ and $i \in \{1, 2, 3\}$, as well as a strategy $t_k$ for each variable $k$. The first matrix contains the payoffs for the row-player, the second one column-player's payoffs.

\rotatebox{90}{\begin{minipage}{\textheight}
$$\begin{array}{c|ccccccc}
 & t_c^1 & t_c^2 & t_c^3 & t_c & t_k\\
 \hline
      s    & 0 & 0 & 0 & 2 & 0\\
    s_d    & \begin{cases} 1 & c = d \\ 0 & \textnormal{else}\end{cases} & \begin{cases} 1 & c = d \\ 0 & \textnormal{else}\end{cases} & \begin{cases} 1 & c = d \\ 0 & \textnormal{else}\end{cases} & \begin{cases} 1 & c = d \\ 0 & \textnormal{else}\end{cases} & 0\\
    s_{l+} & 0 & 0 & 0 & 0 & \begin{cases} 1 & l = k \\ 0 & \textnormal{else}\end{cases}\\
    s_{l-} & 0 & 0 & 0 & 0 & \begin{cases} 1 & l = k \\ 0 & \textnormal{else}\end{cases}
\end{array} $$ $$
\begin{array}{c|ccccccc}
 & t_c^1 & t_c^2 & t_c^3 & t_c & t_k\\
 \hline
      s    & 0 & 0 & 0 & 0 & 1 \\
    s_d    & \begin{cases} 2 & c = d \\ 0 & \textnormal{else}\end{cases} & \begin{cases} 2 & c = d \\ 0 & \textnormal{else}\end{cases} & \begin{cases} 2 & c = d \\ 0 & \textnormal{else}\end{cases} & \begin{cases} 1 & c = d \\ 0 & \textnormal{else}\end{cases} & 0\\
    s_{l+} & \begin{cases} 1 & c_2 = l \vee c_3 = l \\ 0 & \textnormal{else}\end{cases} & \begin{cases} 1 & c_1 = l \vee c_3 = l \\ 0 & \textnormal{else}\end{cases} & \begin{cases} 1 & c_1 = l \vee c_2 = l \\ 0 & \textnormal{else}\end{cases} & \begin{cases} 1 & l \in c \\ 0 & \textnormal{else}\end{cases} & \begin{cases} 1 & l = k \\ 0 & \textnormal{else}\end{cases}\\
s_{l-} & \begin{cases} 1 & c_2 = \neg l \vee c_3 = \neg l \\ 0 & \textnormal{else}\end{cases} & \begin{cases} 1 & c_1 = \neg l \vee c_3 = \neg l \\ 0 & \textnormal{else}\end{cases} & \begin{cases} 1 & c_1 = \neg l \vee c_2 = \neg l \\ 0 & \textnormal{else}\end{cases} & \begin{cases} 1 & \neg l \in c \\ 0 & \textnormal{else}\end{cases} & \begin{cases} 1 & l = k \\ 0 & \textnormal{else}\end{cases}
\end{array}$$
\end{minipage}}

We claim that the strategy $s$ is eventually eliminable, if and only if there is a satisfying truth assignment of the original formula. For the first direction, assume that a satisfying truth assignment is given. For each variable $l$, the strategies $s_{l+}$ and $s_{l-}$ weakly dominate each other. Thus, we can eliminate $s_{l+}$, if the variable $l$ is assigned the value \emph{true}, and $s_{l-}$ otherwise.

In the next step we will eliminate all strategies $t_d$. For each clause $d$, one of its literals must be true. Assume that the $i$th literal for some given clause $d$ is true. We claim that $t_d$ is now weakly dominated by $t_d^i$. $t_d^i$ obviously provides better or equal payoff against $s$, all strategies $s_c$ and all strategies $s_{l+}$ where the literal $l$ does not occur in $d$, as well as all strategies $s_{l-}$ where the literal $\neg l$ does not occur in $d$. If the literal $l$ ($\neg l$) does occur in $d$, but not on the $i$th position, both $t_d^i$ and $t_d$ give payoff $1$ against $s_{l+}$ (against $s_{l-}$). If the literal $l$ ($\neg l$) occurs on the $i$th position in $d$, then, by assumption, $l$ is \emph{true} (\emph{false}), thus, the problematic strategy $s_{l+}$ ($s_{l-}$) has been removed in the first step. Thus, we have covered all cases, and shown that $t_d$ is indeed always eliminable, provided that there is a satisfying truth assignment for the formula.

Once all strategies $t_d$ are removed, the strategy $s$ is weakly dominated by every remaining strategy, thus, it can be removed.

For the other direction, we have to show that if $s$ can be removed, there has to be a satisfying truth assignment for the formula. We will assume that the elimination process was stopped immediately after the removal of $s$.

In the first step, we will show that for each variable $l$, only one of the strategies $s_{l+}$ and $s_{l-}$ was eliminated. For that, we observe that as long as $t_l$ is present, the two strategies are the only ones weakly dominating the other one, thus, only one of them can be removed prior to the removal of $t_l$. Now as long as $s$ is present, a strategy $t_l$ might only be weakly dominated by a strategy $t_k$. However, as either $s_{l+}$ or $s_{l-}$ is still present at this point, $t_k$ does not weakly dominate $t_l$ for $l \neq k$. Whether $s_{l+}$ or $s_{l-}$ was eliminated determines the truth value assigned to the variable $l$. If neither was eliminated, the truth value can be chosen arbitrarily.

As $s$ is the only strategy providing row-player with a payoff of $2$ against any strategy $t_d$, all the strategies $t_d$ were eliminated prior to $s$. We claim that $t_d$ must have been weakly dominated by some strategy $t_d^i$. This is obviously true, provided that $s_d$ was not removed previously. Now $s_d$ can only be weakly dominated, if all strategies $t_d^j$ for $j \in \{1, 2, 3\}$ are removed first, which in turn would require removal of $s_d$, showing that $s_d$ will not be eliminated at all. Therefore, there must be an $i$, such that $t_d$ was weakly dominated by $t_d^i$. Assume that the $i$th literal in $d$ was $l$ $(\neg l)$. Then $t_d^i$ gives less payoff to column-player against $s_{l+}$ (against $s_{l-}$) then $t_d$, thus, $s_{l+}$ ($s_{l-}$) was removed first. According to the construction of our truth assignment, this means that the $i$th literal occurring in $d$ is true, so the clause $d$ is also true. As these consideration applied to all clauses, the truth assignment constructed satisfies the formula.
\end{proof}
\end{theorem}

\section{Conclusions}
If one is willing to accept the whole of $P$ as efficiently computable, our results show that both iterated elimination of strictly dominated strategies as well as iterated simultaneous elimination of dominated strategies are valid as efficient solution concepts for arbitrary games, while neither iterated elimination of weakly dominated strategies nor iterated elimination of dominated strategies can be regarded as such. For zero-sum games, however, all considered concepts can be applied in polynomial time, allowing no distinction on the basis of computational efficiency.

Things change drastically if efficient execution on a parallel machine is required: Suddenly, we can only apply iterated elimination of strictly dominated strategies, and only in the case of two different payoffs, or of three different payoffs in a zero-sum game. This might seem a devastating result, suitable as motivation for either abandoning the request for efficient execution on parallel machines, or the concept of iterated elimination of (weakly / strictly) dominated strategies altogether.

Instead, we suggest a new version of iterated elimination of strictly dominated strategies. Before the elimination process is started, all payoffs are compared to some benchmark value: Low values are replaced by $0$, high values by $1$. This can be done in $NL$ easily, even for a wide variety of procedures to determine the benchmark value. For example, the median payoff level could be used. Then, fast elimination becomes possible. It might even be speculated that such a concept is closer to actual human thinking than the original version.

\end{document}